\documentclass{article}
\usepackage{spconf,amsmath,graphicx,url,subfig, cite,booktabs}
\usepackage{graphicx}
\usepackage{amsfonts,amssymb}
\usepackage{xcolor}
\usepackage{lipsum}

\title{An audio-quality-based multi-strategy approach for target speaker extraction in the MISP 2023 Challenge}

\name{
\begin{tabular}{c}
\it Runduo Han$^1$, Xiaopeng Yan$^1$, Weiming Xu$^1$, Pengcheng Guo$^1$, Jiayao Sun$^1$, \\
He Wang$^1$, Quan Lu$^2$, Ning Jiang$^2$, Lei Xie$^{1*}$\thanks{* Corresponding author.}
\end{tabular}
}
\address{
  $^1$Audio, Speech and Language Processing Group (ASLP@NPU), School of Software, \\ Northwestern Polytechnical University, Xi'an, China\\
  $^2$Mashang Consumer Finance Co., Ltd, China\\
  }

\begin{document}
\ninept
\maketitle
\begin{abstract}
\vspace{-5.8pt}
This paper describes our audio-quality-based multi-strategy approach for the audio-visual target speaker extraction (AVTSE) task in the \textbf{M}ulti-modal \textbf{I}nformation based \textbf{S}peech \textbf{P}rocessing (MISP) 2023 Challenge. Specifically, our approach adopts different extraction strategies based on the audio quality, striking a balance between interference removal and speech preservation, which benifits the back-end automatic speech recognition (ASR) systems. Experiments show that our approach achieves a character error rate (CER) of 24.2\% and 33.2\% on the Dev and Eval set, respectively, obtaining the second place in the challenge.
\end{abstract}

\vspace{-4pt}

\vspace{-1.8em}
\section{Introduction}
\label{sec:intro}
\vspace{-0.8em}
The objective of target speaker extraction (TSE) is to extract the speech of a specific speaker from complex acoustic environments, including background noise and multiple speakers interference. Various research has been conducted in this field \cite{wu2023multimodal,ju2023tea}. These methods typically depend on pre-recorded registration audio of the target speaker, a prerequisite that hinders their widespread practical utility \cite{wu2023multimodal}. Moreover, these methods are often evaluated by audio quality metrics, which may lead to \textit{over-enhancement} for the following ASR task \cite{delcroix2018single}.
In response, the MISP2023 challenge \cite{wu2023multimodal} introduces the readily accessible lip movements video data as prior information for the TSE tasks, instead of target speaker registration audio. To ensure the efficacy of the front-end extracted audio for back-end ASR systems, the challenge incorporates a frozen-parameter ASR system \cite{dai2023improving} to evaluate the character error rate (CER) of the processed audio. 

To cope with such audio-visual target speaker extraction (AVTSE) task, we first use DNSMOS OVRL \cite{reddy2022dnsmos} score to categorize the audio into three quality-based groups, subsequently implementing different extraction techniques for each category. Secondly, a multi-channel fusion method \cite{ju2023tea} is used to recover the speech signal lost in previous extraction by GSS \cite{raj2022gpu}, and further perform TSE with an MEASE \cite{wu2023multimodal} network using lip movements video data.
Thirdly, we utilize the DRC-NET model \cite{liu2022drc} for noise reduction, modifying the original spectrum mapping approach in \cite{liu2022drc} to a masking approach, which balances the noise reduction and speech distortion. On the official Eval set, our approach achieves a CER of 33.2\%, securing the second position in the final assessment.

\vspace{-1em}
\section{Approach}
\label{sec:format}

\vspace{-1.3em}
\subsection{Network Architecture}
\vspace{-0.5em}

\begin{figure}[!htbp]
	\centering
	\includegraphics[width=0.7\linewidth]{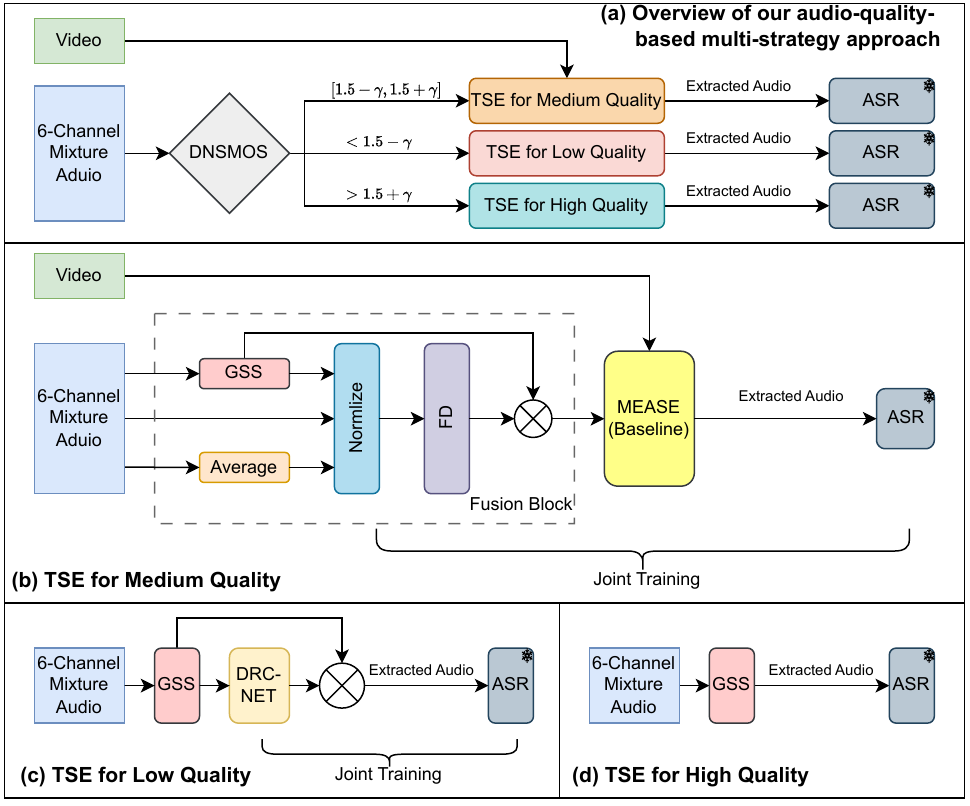}
        \vspace{-4pt}
	\caption{Details of (a) Overview of our audio-quality-based multi-strategy approach; (b)  TSE for Medium Quality audio; (c) TSE for Low Quality audio and (d) TSE for High Quality audio.}
	\label{f2} 
	\vspace{-2em} 
\end{figure}

For audio of varying quality levels, a crucial consideration is whether to prioritize speech preservation with minimal distortion, focus more on noise reduction, or strike a balance between them. Consequently,  we classify audio into three categories based on the DNSMOS OVRL scores: those scoring above $1.5 + \gamma$ are classified as high quality, scoring between $1.5-\gamma$ and $1.5+\gamma$ are considered as medium quality, and any below $1.5-\gamma$ are named low quality. One point five($1.5$) is the DNSMOS OVRL score of the baseline system on Dev set \cite{wu2023multimodal}, which we consider as an intermediate value that differentiates audio quality. The threshold $\gamma$ is a hyperparameter which is tuned from \{0.1, 0.2, 0.3, 0.4, 0.5\} in our experiments, and we find the best results when $\gamma=0.3$. Different processing strategies are then applied to each category, as shown in Fig.\ref{f2}(a). 

\textbf{For high quality audio}, we directly apply the guided source separation (GSS) method \cite{raj2022gpu} as shown in Fig.\ref{f2}(d), which is advantageous for its minimal speech distortion during the extraction of multi-speaker audio. However, for medium or low quality audio, the GSS will mistakenly lose the speech signal of the target speaker and has a poor effect on removing interference. Therefore, some other methods need to be applied to enhance the speech signal after preliminary extraction by GSS. 

\textbf{For medium quality audio}, we add a fusion block for combining multi-channel information and making up for the missing speech signal caused by GSS, and leverage the MEASE network \cite{wu2023multimodal} for further extraction. As illustrated in Fig.\ref{f2}(b), the fusion block combines 8-channel input audio consisting of far-field 6-channel audio, its average, and the GSS processed audio into a single-channel output. 
The 8-channel audio is firstly normalized to match a loudness of -25db. Subsequently, it undergoes a frequency down-sampling (FD) layer to merge magnitude information from multiple channels. The FD layer, structured similarly to \cite{ju2023tea}, includes gated convolutions, cumulative layer normalization (cLN), and PReLU. The output of the FD layer is used as a mask on the averaged magnitude of the far-field 6-channel audio. The masked time-frequency domain audio is then converted back to the time domain and fed into the MEASE network, which uses the target speaker's lip movements video as prior information of the target speaker and further extracts the target speech from the audio. Its structure is similar as the official baseline \cite{wu2023multimodal}, except that the audio and video embedding extraction modules are changed into a 34-layer ResNet structure without pre-training.

\textbf{For low quality audio}, we use the DRC-NET \cite{liu2022drc} network to enhance the single-channel audio after GSS, as illustrated in Fig.\ref{f2}(c), which is good at effectively removing noise and reverberation interference. To minimize speech distortion due to excessive speech enhancement, we modify the original network structure from spectrum mapping to masking, opting for a CRM mask \cite{williamson2015complex} that incorporates phase information. The rest of the network structure remains consistent with that described in \cite{liu2022drc}.
\vspace{-4pt}
\vspace{-0.5em}
\subsection{Training Process}
\vspace{-0.5em}
The model training is divided into two stages, using simulation data for pre-training of the front-end system and followed by the joint training with the back-end ASR system on the real data.

In the first stage, we use the mean square error (MSE) loss $L_{\text{MSE}}$ for the fusion block and MEASE network (i.e. medium quality case). When training the DRC-NET network (i.e. low quality case), the loss function is denoted as $L_{\text{DRC-NET}}$:
\begin{equation}
\mathcal{L}_{\text{DRC-NET}} = \alpha \left\| |\hat{S}| - |S| \right\|_2 + (1 - \alpha)\left\| \text{Mag}(\hat{S}) - \text{Mag}(S) \right\|_2
\end{equation}
where, ${S}$ and $\hat{S}$ denote the target and the estimated spectrum, $\text{Mag}(\cdot)$ is the operation of extracting the magnitude spectrum, and $\lVert \cdot \rVert_2$ is the \textit{L2} norm. $\alpha$ is set to 0.5.

In the second stage, the loss function $L_{\text{ASR}}$ applied for joint training with the back-end ASR system is consistent with that described in \cite{wu2023multimodal}, which is the combination of CTC loss and the CE loss.

\vspace{-1em}
\section{Experiments}
\vspace{-1em}
\label{sec:pagestyle}

\subsection{Datasets}
\vspace{-0.5em}

We utilize the MISP2023 Challenge dataset for the experiments. The simulation method for the far-field 6-channel data is similar to that described in \cite{wu2023multimodal}. During the fist training stage, we perform dynamic noise and speech mixing, randomly adding noise ranging from -10db to +20db to the clean audio and combining it with other near-field speaker audios. Then, reverberation is added according to room dimensions to simulate far-field 6-channel audio. The audio data used for the second training stage, joint training with the back-end ASR model, is real-scene far-field 6-channel audio. All lip movements video data used as prior information in speech extraction is recorded with mid-field cameras. For model inference, both the Dev set and Eval set are from the official MISP 2023 challenge dataset.

\vspace{-1.3em}
\subsection{Experiment Setup}
\vspace{-0.5em}
For the short-time fourier transform (STFT), we use a window length of 32ms and a hop size of 10ms, with an STFT length of 512. The training is conducted using the Adam optimizer, with an initial learning rate set at 0.001. The learning rate will be halved if the validation loss has no decrease for 3 epochs.


In the fusion block's FD layer, the gated convolution has a kernel size and stride of (2, 3) and (1, 2) respectively, corresponding to the time and frequency axes. In the DRC-NET, the encoder and decoder each contain 4 layers of DRC blocks and DeDRC blocks. Within each block, kernel size and stride of gated convolution are (1, 5) and (1, 2) respectively, and the channel dimensions are \{2, 32, 32, 64\}.

\vspace{-0.7em}
\begin{table}[htbp]
\centering
\caption{Detailed CER (\%) and DNSMOS OVRL results of different approaches on the official Dev set.}
\vspace{-5pt}
\begin{tabular}{lcccc}
\hline
\textbf{System} & \textbf{CER \% $\downarrow$} & \textbf{DNSMOS $\uparrow$} \\ \hline
GSS & 26.4 & 1.35 \\ \hline
MEASE (Baseline) & 26.3 & 1.50 \\ \hline
Fusion + MEASE & 25.6 & 1.42 \\ \hline
DRC-NET & 27.8 & \textbf{1.86} \\ \hline
Our Approach (NPU-MSXF) & \textbf{24.2} & 1.47 \\ \hline
\end{tabular}
\label{tab:my_label}
\end{table}

\vspace{-2.3em}
\begin{table}[htbp]
\centering
\caption{Detailed CER (\%) and DNSMOS OVRL results of different approaches on the official Eval set.}
\vspace{-5pt}
\begin{tabular}{lcccc}
\hline
\textbf{System} & \textbf{CER \% $\downarrow$} & \textbf{DNSMOS $\uparrow$} \\ \hline
GSS & 37.6 & 1.31 \\ \hline
MEASE (Baseline) & 36.1 & \textbf{1.42} \\ \hline
Our Approach (NPU-MSXF) & \textbf{33.2} & 1.41 \\ \hline
\end{tabular}
\label{tab:my_label}
\end{table}
\vspace{-2em}

\subsection{Results}
\vspace{-0.6em}

Ablation experiments on the Dev set, detailed in Table 1, evaluate each component's efficacy. Results from DNSMOS OVRL score indicate that post-GSS neural networks integration reduces noise, but at the cost of introducing some distortion to speech. Hence, employing varied extraction strategies for different audio qualities is crucial, leading to the lowest CER in Dev set. The fusion block, when combined with the MEASE network, achieves an enhanced CER compared to using the MEASE network alone, evidencing the fusion block's effectiveness in integrating multi-channel information. In the final evaluation, presented in Table 2, our approach significantly outperforms GSS and MEASE models in CER, securing the second place in the MISP2023 challenge. It is worth noting that a higher DNSMOS OVRL score does not always align with a lower CER, highlighting that for front-end enhancement models serving back-end ASR systems. In other words, when serving back-end ASR systems, audio quality metrics cannot be used alone for evaluation.

\footnotesize
\bibliographystyle{IEEEbib}
\let\oldbibliography\thebibliography 
\vspace{-5pt}
\renewcommand{\thebibliography}[1]{ 
  \oldbibliography{#1}
  \setlength{\itemsep}{-1pt} 
}
\bibliography{strings,refs}

\end{document}